\documentclass[prl,twocolumn,showpacs,amsmath,amssymb]{revtex4}
\usepackage{dcolumn}
\usepackage{bm}
\usepackage{graphicx}
\usepackage{color}

\begin{document}

\title{Finite-size effect on evolution of Griffiths phase in manganite nanoparticles}

\author{A. K. Pramanik}\email{akpramanik@mail.jnu.ac.in}\altaffiliation{Present address: School of Physical Sciences, Jawaharlal Nehru University, New Delhi - 110067, India.}\affiliation{UGC-DAE Consortium for Scientific Research, University Campus, Khandwa Road, Indore - 452001, M.P, India.}
\author{A. Banerjee}\affiliation{UGC-DAE Consortium for Scientific Research, University Campus, Khandwa Road, Indore - 452001, M.P, India.}

\begin{abstract}
The finite-size effect on the evolution of Griffiths phase (GP) is studied using nanoparticles of half-doped manganite Pr$_{0.5}$Sr$_{0.5}$MnO$_{3}$ with different average particle sizes but having similar structural parameters. All the samples exhibit pronounced GP behavior. With reducing particle size, the Griffiths temperature remains almost unchanged but the characteristic critical temperature $T_c^R$ decreases and the GP properties are strengthened. It is noteworthy that the shift of $T_c^R$ follows finite-size scaling with the particle size revealing an exotic interplay between the GP properties and the sample dimension. This reinforces an earlier proposal of length-scale related evolution of GP.   
\end{abstract}


\maketitle
Nanosized materials continue to attract large amount of scientific interest. With this reduced dimensions, the physical properties are mainly governed by two factors viz., surface and finite-size effect \cite{batlle}. While the former is realized due to broken translational symmetry and magnetic exchange bonds at the surface which leads to surface spin disorder and frustration the later originates due to hindrance of divergence of long-range correlation length at the system boundaries. The interplay between these two effects often manifest many exotic phenomena such as, modification of transition temperature and saturation magnetization, weakening of antiferromagnetic (AFM) state and emergence of ferromagnetic (FM) state, appearance of superparamagnetic (SPM) behavior and glassy dynamics, exchange bias, etc \cite{tang,dutta,rivadulla,biswas,sako,rao,zhang,psmo-nano,kodama,huang}.

In this manuscript, we have studied the effect of finite-size on the evolution of Griffiths phase (GP) behavior \cite{griffiths}.  The GP arises due to a phase inhomogeneity above the long-range FM ordering temperature $T_c$, and the quenched disorder serves as main ingredient for its occurrence. Originally, the GP was described for a Ising ferromagnet considering that if a system is randomly diluted either creating vacancy or replacing the magnetic atoms with the nonmagnetic ones then there is a distribution of nearest-neighbor exchange bonds with the values $J$ and 0 having probability $p$ and 1 - $p$, respectively. In this scenario, the transition temperature of the diluted system $T_c(p)$ becomes less than that for the pure one i.e., $T_c(p=1)$ which is also recognized as the Griffiths temperature $T_G$. For $p < p_c$ ($p_c$, the percolation threshold), the system does not develop any long-range ordering. The temperature ($T$) regime between $T_c(p)$ and $T_G$ is, however, quite interesting where the system neither exhibit pure paramagnetic (PM) nor long-range FM ordering. Instead, there exists spatially distributed small clusters of different sizes with local FM ordering. This leads to a situation called `Griffiths phase' where the magnetization happens to become nonanalytic. Usually, Griffiths singularity is characterized by the susceptibility ($\chi$) exponent $\lambda$, ($0 < \lambda \leq 1$) with following relation \cite{neto},

\begin{eqnarray}
	\chi^{-1} \propto (T - T_c^R)^{1-\lambda}
\end{eqnarray}

It is clear that the power law behavior in Eq. 1 is a modified Curie-Weiss (CW) law where the exponent $\lambda$ quantifies a deviation from CW behavior due to formation of magnetic clusters in PM state. As $T_c$ is approached from above, more number of clusters achieve FM ordering and the bulk susceptibility of system tend to diverge at $T_c^R$ (usually $>$ $T_c$) which is regarded as the critical temperature of random ferromagnet \cite{salamon,venka,william-prl,william,psmo-gp}. Recently, there have been several experimental studies to identify the $T_c^R$, yet the underlying role of $T_c^R$ in modulating GP behavior is not clear. However, the divergence of bulk susceptibility at temperature $T_c^R$ (Eq. 1) higher than $T_c$ implies that even if the sample does not attain long-range FM ordering, perhaps there develops inter-cluster correlation of magnetic fluctuations. If this correlation exists, the question raises about its typical length scale and its critical nature at $T_c^R$. In this regard, imposing a physical constraint on the thermal evolution of correlation length will be a crucial step which will also probably help to understand how a system with GP behavior thermally evolves from PM to FM phase, as recently new scaling relations have been proposed near the GP-FM phase transition \cite{salamon-crit}.
        
In the last several years, the possible existence of GP has been shown in many kind of materials including transition metal oxides (TMOs) \cite{salamon,deisenhofer,rama,venka,william-prl,william,psmo-gp,shimada,galitski,magen,sampath,pecharsky}. The TMOs have in-built natural disorder contributed by random size distribution of cations. The most recent studies on GP behaviour have focused on the role of structural disorder coming from variation in tri- and di-valent ions which eventually translates to a case of `bond disorder' \cite{salamon,deisenhofer,rama,venka,william-prl,william}. On the other hand, the evolution of GP has also been investigated for `site disorder' introducing chemical substitution at Mn-site \cite{psmo-gp}. Similarly, in nanoparticles the GP behavior has been observed for several doped manganites, mostly due to size induced weakening of charge-ordered antiferromagntic interaction present in their bulk materials \cite{lu,zhang-np,zhou}, but the role of finite-size effect on the development of GP is yet to be investigated. 

In present study, we have used nanoparticles of half-doped manganite Pr$_{0.5}$Sr$_{0.5}$MnO$_{3}$ (PSMO) as a model system to understand the size controlled evolution of magnetic fluctuations. The typical size of FM clusters above $T_c$ in TMOs is about 1 - 2 nm \cite{teresa,he}, and our nanoparticles are sufficiently large (Table 1) to support such clusters. The bulk PSMO shows 2nd order PM to FM transition followed by a 1st order FM to AFM transition with lowering in temperature \cite{tomioka,alokjpcm,psmo-ps,psmo-crit}. Recently, we have shown strong GP behavior in bulk PSMO where it is further strengthened with the substitution of nonmagnetic Ga at Mn-site, which has been ascribed due to change in length scale of FM clusters \cite{psmo-gp}. In fact, the present tailor-made nanoparticles with different sizes offer an ideal playground to test this length scale induced modification in GP properties. These nanoparticles have shown some interesting features such as, SPM behaviour, a dipolar-type interparticle interaction, crossover in critical lines on field-temperature plane, exchange bias effect, memory effect, etc \cite{psmo-nano,psmo-conf}. In present study, we show a pronounced GP behavior in these nanoparticles of PSMO at high-$T$ across PM to FM phase transition. With reducing the particle size, though $T_G$ remains almost constant but the $T_c^R$ decreases and the susceptibility exponent $\lambda$ increases which imply an enhancement of both GP regime as well as its strength. It is further noteworthy that modification of $T_c^R$ follows a finite-size scaling behavior exhibiting a prominent role of finite-size on the evolution of GP.

\begin{table}
\caption{\label{tab:table 1} The particle sizes, critical temperatures and the inverse susceptibility exponents ($\lambda$) for Pr$_{0.5}$Sr$_{0.5}$MnO$_{3}$ nanoparticles.}
\begin{ruledtabular}
\begin{tabular}{ccccc}
Samples &N600 &N650 &N700\\
\hline
D$_{XRD}$ (nm) &15.7 &17.3 & 19.1\\
D$_{TEM}$ (nm) &15.7 &19.2 & 26.6\\
D$_{Mag}$ (nm) &16.5 &19.2 & 20.9\\
T$_G$ (K) &316(2) &318(2) &316(2)\\
T$_c^R$ (K) &260.1 &274.1 &280.1\\
$\lambda_{PM}$ &0.026(8) &0.02(9) &0.016(6)\\
$\lambda_{GP}$ &0.80(3) &0.74(4) &0.64(6)\\
\end{tabular}
\end{ruledtabular}
\end{table}  

Nanocrystalline samples of PSMO (three batches designated as N600, N650 and N700) are prepared using a chemical pyrophoric method, and details are given elsewhere \cite{psmo-nano}. The particle sizes as obtained from x-ray diffraction ($D_{XRD}$), transmission electron microscope ($D_{TEM}$) and magnetization data ($D_{Mag}$) are given in Table 1. All the samples are found to be chemically pure, and without significant mismatch in structural parameters due to size variation. The dc magnetic measurements are performed with a vibrating sample magnetometer (PPMS, Quantum Design) and ac susceptibility is measured with a home-built instrument \cite{ashnarsi}.

\begin{figure}
	\centering
		\includegraphics[width=8cm]{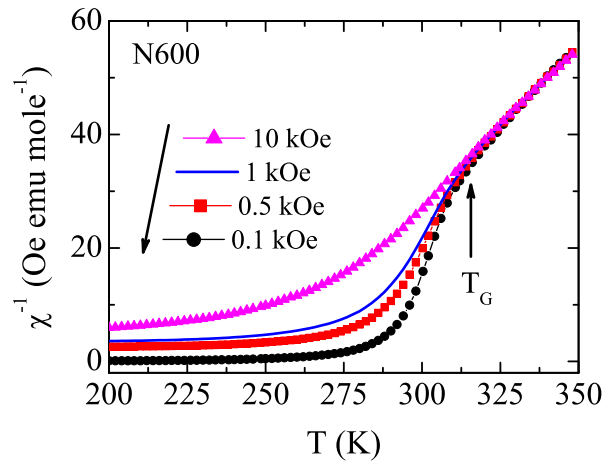}
	\caption{(Color online) Temperature variation in inverse dc magnetic susceptibility ($\chi^{-1}$) measured in 0.1, 0.5, 1 and 10 kOe are plotted for Pr$_{0.5}$Sr$_{0.5}$MnO$_{3}$ nanoparticles (N600). The vertical up arrow marks the Griffiths temperature $T_G$.}
	\label{fig:Fig1}
\end{figure}

The GP is characterized as divergence in susceptibility which implies that an inverse susceptibility ($\chi^{-1}$) would exhibit a sharp downturn with decreasing temperature \cite{neto}. Fig. 1 shows temperature dependence of $\chi^{-1}$ deduced from dc magnetization measured in different magnetic field ($H$) for representative N600 material. At high temperatures, $\chi^{-1}(T)$ exhibits a linear dependence for all the measuring field following the CW behavior. As temperature decreases, the $\chi^{-1}$ deviates from linearity showing a downturn which is typical to GP behavior. Note, that similar downturn in $\chi^{-1}$ has also been observed in low field ac susceptibility data (not shown). At lowest measuring field of 0.1 kOe the downturn is reasonably sharp while with increasing field (0.5, 1 and 10 kOe) the sharpness of downturn is reduced. This can be understood as in higher fields the magnetic response from the PM matrix becomes substantial to dominate over the embedded FM clusters, hence driving toward a linear behavior in $\chi^{-1}(T)$. The observed magnetic behavior in Fig. 1 confirms GP in present naoparticles as the mere presence of magnetic clusters above $T_c$ does not necessarily lead to Griffiths singularity without a divergence in susceptibility \cite{teresa,he}.  
    
The presence of Griffiths singularity in all nanoparticles has been confirmed from $\chi^{-1}$ vs $T$ measured in 0.1 kOe (Fig. 2). All the samples exhibit a sharp downturn in $\chi^{-1}$, however, in high-$T$ PM regime the $\chi^{-1}$ monotonically decreases with the particle size. This implies a relatively reduced PM moment in smaller size particles. Indeed, we have estimated the effective PM moment $\mu_{eff}$ as 3.62, 3.70 and 4.38 $\mu_B$/f.u. for N600, N650 and N700, respectively while the expected spin-only value ($g\sqrt{S(S+1)}$) of PSMO is about 4.38 $\mu_B$/f.u. \cite{psmo-nano}. This low value of $\mu_{eff}$ is not though related to the stoichiometry of sample as Iodometric titration reveals $Mn^{3+}$/$Mn^{4+}$ ratio is close to 1 for all the samples. However, the prominent surface disorder in smaller size particles may cause such low $\mu_{eff}$, as similarly observed for low-$T$ moment \cite{psmo-nano}.

\begin{figure}
	\centering
		\includegraphics[width=8cm]{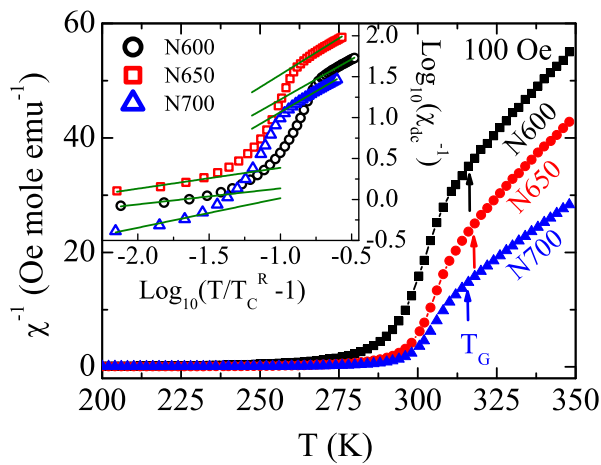}
	\caption{The inverse magnetic susceptibility ($\chi^{-1}$) is plotted as a function of temperature for all  the Pr$_{0.5}$Sr$_{0.5}$MnO$_{3}$ nanoparticles. The vertical arrows mark the $T_G$. In inset, the susceptibility data are plotted against temperature following Eq. 1 in double logarithmic scale. The data for N650 is vertically shifted by 0.35 for clarity. The straight lines are due to straight line fitting of Eq. 1.}
	\label{fig:Fig2}
\end{figure}

From the onset of downturn in $\chi^{-1}(T)$ we have estimated the $T_G$ for all the samples (see Table 1) \cite{magen}. With marked observation, the $T_G$ remains almost constant with varying particle size, even values are very close to $T_G$ (315 K) for the bulk PSMO material \cite{psmo-gp}. While the long-range magnetic ordering temperatures are susceptible to the size variation of materials but the $T_G$ which is considered to be the ordering temperature of pure disorder-free material proves to be very robust. Further, we have characterized the GP utilizing Eq. 1. The inset of Fig. 2 shows log$_{10}$-log$_{10}$ plot of $\chi^{-1}$ vs $T/T_c^R$ - 1, and from the slope of straight line fitting (Eq. 1) we have obtained the exponent $\lambda$. Following previous discussion, the $\lambda$ in Eq. 1 presents a means to measure the strength of GP as in pure PM case $\lambda$ $\approx$ 0 and the system follows the original CW behavior. However, the determination of proper $\lambda$ is very sensitive to right value of $T_c^R$. We have followed a rigorous method prescribed in Ref. \onlinecite{psmo-gp} to determine the $T_c^R$, and subsequently $\lambda$ (Table 1). The very negligible $\lambda_{PM}$ in PM regime (above $T_G$)  implies that obtained $T_c^R$ are authentic. The obtained $\lambda_{GP}$ in GP regime are quite comparable with other TMOs \cite{salamon,william-prl,william,psmo-gp,magen,sampath,pecharsky}. We observe that $\lambda_{GP}$ increases and $T_c^R$ decreases with reducing the particle size.

\begin{figure}
	\centering
		\includegraphics[width=8cm]{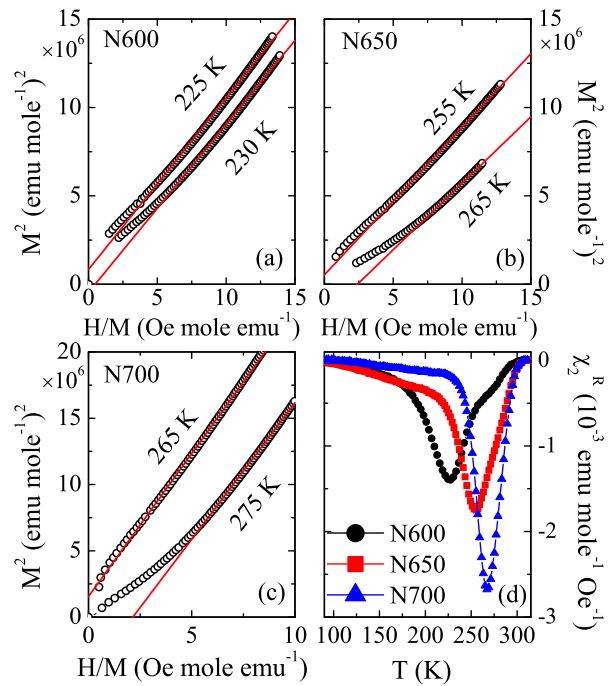}
	\caption{Magnetic isotherms collected at different temperatures are plotted as $M^2$ vs $H/M$ (Arrott plot) for (a) N600, (b) N650 and (c) N700 nanoparticles of Pr$_{0.5}$Sr$_{0.5}$MnO$_{3}$. (d) shows temperature dependence of real part of second order ac susceptibility $\chi_2^R$ measured in frequency 731 Hz and ac field 2.5 Oe.}
	\label{fig:Fig3}
\end{figure}
   
 In the GP regime there are clusters with short-range FM ordering but the system as a whole does not exhibit spontaneous magnetization $M_s$ \cite{vojta}. This has been examined using 2nd order ac susceptibility ($\chi_2$) and Arrott plot, respectively. For Arrott plot, the magnetic isotherm $M(H)$ is plotted in the form of $M^2$ vs $H/M$, and the positive intercept on $M^2$ axis due to an extrapolation of straight line fitting in high field regime gives $M_s$ \cite{arrott}. In Figs. 3a, 3b and 3c we present few representative Arrott plots around $T_c$ for N600, N650 and N700, respectively. The figures clearly show that all the samples develop $M_s$ at temperatures ($T_c$) much lower than $T_c^R$ (Table 1). This confirms that in GP regime there is no $M_s$ though there are ferromagnetically ordered small clusters. To further elucidate this issue we have measured $\chi_2$ which is only evident due to FM spin correlations and diverges with the appearance of $M_s$ at $T_c$ \cite{psmo-ps,psmo-gp}. Henceforth, the $\chi_2$ is not expected in PM state above $T_c$. The temperature dependence of $\chi_2^R$ (real part of $\chi_2$ measured in 731 Hz and 2.5 Oe ac field) in Fig. 3d shows a finite appearance at low temperature, however, it exhibits a pronounced peak around 227.5, 255.3 and 267.2 K for N600, N650 and N700, respectively and these temperatures are in agreement with the appearance of $M_s$ as observed in Fig. 3a, 3b and 3c. Interestingly, a finite value of $\chi_2^R$ is seen to be present above the peak ($T_c$) way up to temperature very close to $T_G$ for all the samples. These observations experimentally verifies that in GP regime the system does not hold $M_s$ related to long-range FM ordering but there exists FM spin fluctuations.
            
Interestingly, the significant particle size dependence of characteristic temperature $T_c^R$ (Table 1) can be nicely described following finite-size scaling theory \cite{batlle,fisher},

\begin{eqnarray}
	1 - \frac{T_c^R(D)}{T_c^R(\infty)} = \pm \left(\frac{D}{D_0}\right)^{-1/\nu}
\end{eqnarray}

\begin{figure}
	\centering
		\includegraphics[width=8cm]{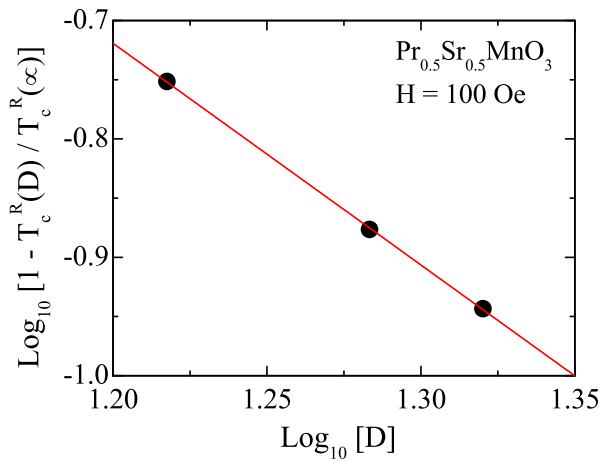}
	\caption{The scaling behavior of reduced temperature of $T_c^R$ vs particle size is shown in double logarithmic scale. The straight line is fitting following Eq. 2.}
	\label{fig:Fig4}
\end{figure} 
  
where $D$ is the diameter (or size) of particles, $T_c^R(\infty)$ is related temperature for infinite size system, $D_0$ is the characteristic length scale on microscopic level and $\nu$ is the exponent for correlation length. We have taken $T_c^R$ and $D_{mag}$ from Table 1, and we are quite justified in taking $T_c^R(\infty)$ = $T_G$ (316 K) as for infinite size particles the $\chi$ is expected to diverge at $T_G$. The data plotted in Fig. 4 following Eq. 2 show a nice finite-size scaling behaviour, and the fitting yields the exponent $\nu$ = 0.535(5) and $D_0$ = 6.54(2) nm. The similar nice finite-size scaling has also been obtained for $T_c$ (taking temperature where $\chi_2^R(T)$ shows peak in Fig. 3d and $T_c(\infty)$ = $T_G$) with $\nu$ = 0.398(8) and $D_0$ = 9.94(3) nm (not shown). While the theoretical predictions for $\nu$ vary based upon different magnetic models such as, between 0.65(7) - 0.733(20) for Heisenberg model and 0.5 for the mean-field model \cite{kaul,stanley} the presently obtained $\nu$ at $T_c^R$ and $T_c$ does not match with the established magnetic models. This is significant because the critical exponents for $M_s$ and $\chi$ in bulk PSMO do not also agree with the established universality classes \cite{psmo-crit}. Moreover, the $\nu$ $\neq$ 1 in present case imply that modification of $T_c^R$ in Fig. 4 is not due to a pure surface effect as the surface/volume $\propto$ $D^{-1}$.

The shift of transition temperature following finite-size scaling theory has previously been observed in case of thin films as well as nanoparticles \cite{farle,li,tang,wang,shu}. While the reasonable size mismatch between Pr$^{3+}$ and Sr$^{2+}$ ions and the deviation from cubic perovskite structure causes a structural disorder which perhaps contributes to GP behavior in Pr$_{0.5}$Sr$_{0.5}$MnO$_{3}$, it is notable that the structural parameters and the ionic concentration do not change significantly with the size variation \cite{psmo-nano}. This is substantiated by the fact that the $T_G$ remains unaltered for the nanoparticles, which is also the $T_G$ of the bulk PSMO \cite{psmo-gp}. In this respect, the evolution of GP and such scaling of $T_c^R$ with particle size are significant. The basic comprehension of $T_c^R$ is a critical temperature where $\chi$ diverges (Eq. 1). Therefore, size induced modification of $T_c^R$ following finite-size scaling in Fig. 4 underlines the fact that there exists correlation of magnetic fluctuations among the clusters above $T_c$ which diverges at $T_c^R$. This divergence is, however, hindered by the size of particles, and further substantiated by the finite-size scaling of $T_c$. The present observations are crucial regarding the GP-PM phase transition while there is ongoing debate about the experimental evidence for the GP behavior. Moreover, an obeying of scaling behavior for all the investigated particles means that the typical magnetic correlation length in this material is larger than the maximum size of particle i.e., N700. The striking effect of particle size is also realized with an increase value of $\lambda_{GP}$ which implies an enhancement of GP property in smaller size materials \cite{psmo-gp}. Nonetheless, the present study provides the first experimental realization of finite-size scaling effect on GP behavior. We hope that theoretical investigations will be extended to comprehend this exotic interplay between the GP behavior and finite-size effect.

In conclusion, to understand the finite-size effect on the GP behavior the size dependent nanoparticles of half-doped manganite Pr$_{0.5}$Sr$_{0.5}$MnO$_{3}$ with similar structural parameters and ionic concentration are prepared. A pronounced Griffiths singularity is observed which has been characterized with FM spin fluctuations above $T_c$ but without spontaneous magnetization. The Griffiths temperature does not modify, but the strength of GP increases and the characteristic GP temperature $T_c^R$ decreases with reducing particle size. The shift of $T_c^R$ nicely follows the finite-size scaling theory. This shows presence of inter-cluster correlation of magnetic fluctuation which becomes critical at $T_c^R$. This study brings out an interesting interplay between finite-size effect and GP properties, and clearly shows the effect of length-scale on the evolution of GP as postulated earlier.

AKP acknowledges CSIR, India for financial support.

\end{document}